\documentstyle[12pt]{article}

\begin{document}
\hoffset = -1truecm
\voffset = -2truecm
\begin{centering}
{\huge\bf System and Its Uncertainty Quanta: Theory of General System (I)}\\
\vskip .3cm
Zhen Wang\\
Physics Department, LiaoNing Normal University, Dalian 116029, P.R.China \\
\vskip .5cm
{\bf Abstract}\\
\end{centering}
\vskip .2cm
The concept of uncertainty quanta for a  general
system  is  introduced  and applied to some
important problems in physics  and  mathematics.
EPR  paradox gives new clue to the further
understanding of particle correlation which turns
out to be the nature of this world.
Randomness  in quantum mechanics, statistical
physics and chaos is integrated. A picture for a new kind of
mathematics is put forward.\\
\vskip .5cm
\noindent {\large\bf I. Introduction}\\
\vskip .3cm
 Our knowledge about the world has been greatly deepened
 with the developments in modern science. People now
 can have a  very  deep  understanding  of  many phenomena
 in varieties of areas. As the  basic  science,
physics  has  seen tremendous changes during
this century. First the theory  of  relativity  and quantum
mechanics; then cross-disciplines
marked  with  the  researches  for complexity.
All these great breakthroughs of human knowledge
are the proudest achievements
in this century. But it
should
also be noted that  neither  can present physics provide an
integrated and unified theoretical frame for
other subjects, such as mathematics, astronomy,
biology and even social sciences,  nor can it get rid of
the haunting of quite a few perplexing problems
within its own area. Progress in cross-disciplines
and frontier sciences  in  recent decades has given us a
lot of inspiration. We can not help suspecting
whether there is a common and tremendous truth lying
under  the  minefield  of  the difficulties
in physics and other subjects. This is the first
thesis  of  my series of work, which aims to get
an  unified  understanding  of  some  basic problems
in physics, mathematics and philosophy.

Science comes from experience. But we
know that at any time of human
history, human experience is always limited,
even not worth  mentioning  relative  to the vast
universe. Science is the abstraction of the
common  part  of  human experience and can not
include the experience of every individuals.  We shall
point out in our work that the differences between
individuals are  not  only significant but also endless.
Thus on one hand, it would be difficult to  get further  understanding
about  the
principles  of  nature, if  we  do
not differentiate individual experience
and common human experience. On the other hand,
scientific induction is always incomplete. Thus
man should not deny a special existence because of the  lack
of  corresponding  experience.   The prevailing
scientific method is to prove true or prove false
within  present theoretical frame. Apparently the
conclusions got in this way would be  under the
influence of the limit of the frame. So all the
conclusions  are
relative and should not be extended wantonly.
In fact, in  a  theoretical  frame  with limitation,
there must be things which can neither  be  proved
true  nor  be proved false. They should not be regarded
as fabricated imaginations, but  as the key to break
through the limitation.  It  is  a  serious  defect
in  our epistemology that such obvious truth is often
ignored. In this paper, I shall put forward the concepts
of general system, the environment  and its uncertainty
quanta. We shall also
use these
concepts to discuss some important problems in physics
and mathematics. The further meaning of these  concepts
will  be discussed in detail in my third paper "Quantum Cosmology".
\vskip .5cm
\noindent {\large\bf II. Uncertainty Quanta of A General System}\\

        1.  System

Anything that can be viewed as a whole may be
regarded as a system. Simple as a single
particle or complex as the whole universe, these
all can be regarded as systems.  Each system has its
own environment and  has  its  own  special
synergistic function relative to the environment.
You may also  say, a  system fixes its environment
with  its  synergistic  function. The  level  of  this
synergistic function depends on the inner order status
of the system. The more ordered the system,
the stronger its synergistic function. To some  extent,  the
defining of a system is arbitrary. But a system is
usually  inconvenient for discussion if
there are different kinds of synergistic  functions
within the system. For example, in a system consisting of
a man and  a  chair,   the synergistic functions within the
man and between the man and  the  chair  are apparently
different. Therefore, such defining of system is not
helpful  in our discussion.

        2.  Synergistic Function and Environment

The synergistic function is the nature and also the way
of being of a system. It's a mark of the order
status of a system. The  synergistic  function  has two
forms  or  works  in  two  ways.   One  is  the  capacity
to  sense  and differentiate the environmental changes, to
respond to the influence  of  the environment, which we call
 responding capacity. The other is the  ability  to exert
 influence to select or fix the environment,  which  we
 call  selecting ability. A system is
more difficult to be  affected  if  it  has  a  stronger
selecting ability, and so it is weaker in its responding
capacity.  They  are the two sides of one coin. Obviously,
a  more  ordered  system  has  stronger selecting ability
and weaker responding capacity. If  only  electro-magnetic
interaction involved, an electron has only one kind of
synergistic function.  It can only respond to and select
environment with this kind of function.   A more
sophisticated system may have  stronger
selecting  ability
and  weaker responding capacity, but generally its
synergistic function is always limited. Thus for any
system, there must be some parts in its environment
which  it  can not select but only to be influenced.
The part of the environment to which the system can
apply its selecting ability is called the inner
environment of the system. Unless specially mentioned,
we usually refer to the inner environment simply as
environment. The part of the environment which the
system can  not fix but only have to respond
to is called the outer
environment of the system. A more ordered system
has a stronger  selecting   ability,   and  so  a
more abundant inner environment. The inevitable
existence of defect or  limitation in a common
system corresponds to its outer environment.
System and  its  environment  compose all objects in our
research, and we can say nothing about things  apart  from these.

        3.  The Uncertainty Quanta of a System

The synergistic function of a system is usually limited.
Therefore
there must be some uncertainty in its environment.
We define  three  uncertainty  quanta  to represent
the uncertainty in the environment of a system. As
will be seen, these uncertainty quanta also reveal
some kind of quantum nature
for the system.

\begin{tabbing}
11111111111111111111\=22222222222222222222\=333333333\=\kill\\
\>{mass quantum:}\>{$m_{q}$}\\
\>{space quantum:}\>{$l_{q}$}\\
\>{time quantum:}\>{$t_{q}$}\\
\end{tabbing}

The mass quantum $m_{q}$ is the smallest mass unit that the
system  can  recognize. According to the relationship
between mass  and energy revealed in Einstein's
theory of relativity, we may also take the mass
quantum $m_{q}$ to be the mark of the selecting
ability of the system. The  smaller  the  mass
quantum  of  a system is, the more accurately and further
it can fix its environment, thus the stronger its
selecting ability is. The Space quantum  $l_{q}$
is  the  shortest length for the system,
within which all points
are  absolutely  equivalent.  Viewed from the angle
of relationship between different systems,  $l_{q}$
reveals the degrees  of  quantum  nature  of  space
in  the  system  and  its environment. The time quantum
$t_{q}$ is the shortest time interval for the system.
It is the basic constructing brick of time. Time interval
that  is  shorter than  $t_{q}$ is meaningless to the system.
These  uncertainty  quanta  may  change their values in
the evolution of the system. In extreme cases,
they  may  be zero or
infinite.
 The more ordered a system,  or the  higher  its
 synergistic level, the smaller its mass and time quanta,
 and the larger its space quantum. This means that the
 selecting  ability  as  well  as  the  quantum character
 of the system have been enhanced, the clock has been
 slowed  down.  Here, the changes of clock have nothing
 to do  with  the  special  theory  of relativity. They
 are  independent  time  transformations. The  difference
  in clocks of two systems with no relativistic effect is
 determined
 by  the  two time quanta.

The speed of light (or the maximum speed) in the system
and  its  environment can be expressed descriptively as
(See a following chapter "Consideration  on Mathematics" )

$$c =  {l_{q}\over{t_{q}}}\eqno(1)$$

\noindent Different systems may have different space quanta
and time quanta, so  it  is natural there may be different
light speeds for different systems. The inside of $m_{q}$
and  $t_{q}$ and the outside of  $l_{q}$ represent the
limitation of the system in its synergistic function.
The uncertainty quanta are  characteristic  nature of a system.
Some special systems in our study may have
uncertainty  quanta other than mass, space and time
quanta. They may also have only  one  or  two uncertainty quanta in
its  one  or  two  special properties.

       4.  Self-centered System

When a part of the environment can not be fixed
by the system,   but  only affects the system, then it  is  in
the  outer  environment  of  the  system according to
our definition. It reveals the
limitation  in  the  synergistic level of the system. A specific
environment belongs to a specific system, and each system
has its own special
environment, although there might be  common or similar
part of environment for different systems.
So in  principle,  people can not talk about the
environment  of  others
who  are  also  part  of  the environment of
the subject. The division of system and environment
endows  a special role to what is defined to be the
system: it  becomes  the  subject.  There is only one
system, everything else are all things in the environment.
The states of a system and its environment on a time quantum
corresponds  to each other. When we regard ourselves  as
systems,   then  the  corresponding relation is the source
of the onlyness in  our  world.   The  continuous  and infinite
correspondence between system and its environment is  the
evolution of the system and its environment. Therefore, in
this  sense,   there  is  no evolution that belongs only to
system or only to the environment. Evolution refers to the
changes of relationship between system and its environment.
The limitation in  the  synergistic  function  results  in
uncertainty  in  the evolution to the system. Thus we see
that the uncertainty in the  environment is caused by the
system's  own  limitation.   When
the  system  applies  its selecting ability, the state
of the system determines its environment to  the degree
marked by its uncertainty quanta.

The physics property of the environment  is  determined
by  the  synergistic function of the observer system. A
constant light speed is a basic presumption in the theory
of special relativity. It was introduced as a law of
experience. In my opinion, light travels at constant
speed because it is measured by the same observer,
according to same theory, using same kind  of  device,
with similar method and in like environment. Such
presumption  may  describe  the inner environment
of mankind
quite well,
but  it  may  also  turn  to  be  a obstacle for a
deeper theoretical insight. I firmly believe that
behind  each law of experience in nature lies something
deeper,   more  natural  and  more harmonic.

\noindent {\large\bf III.   Discussion for EPR Paradox}\\

After quantum mechanics was
put forward, people were
not satisfied  with  its probability character. A group
of physicists headed by Einstein insisted that physics
should be deterministic in principle, rather than  mere
probability description. That gave rise to the famous
debate between Einstein and  Bohr.  In 1935, Einstein,
Podolsky and Rosen designed a theoretical model in
quantum mechanics which designated later as EPR paradox.
In EPR paradox, a system composed of an electron and a
positron  with their spins polarized respectively in +/-z  directions was
set off collinearly with a photo. Since the total angular
momentum and total spin were constant zero, there would be
instant influence to the other particle if we measure the
spin of one particle according to orthodox quantum mechanics,
no matter  how  far apart they were. This, of course, does
not  comply  with  the  sermon  about constant light speed.
Einstein regarded it as a mark of  incompleteness  of the
theory, and Bohr explained
it with quantum mechanics and pointed out that the cause of the
paradox was a wrong viewpoint of natural philosophy we  had been
used to, i.e., measurement would not
affect the object.

But there was something unusual in the explanation of quantum
mechanics: the two particles were regarded as one object. What
makes us take them as a whole? In my theory the  two  particles
are  considered  as  one  system  in  the environment of  the
observer.   They  compose  a  system  because  they  are
"correlated", and this is in turn because they are created
at same time. This means they are created within one time
uncertainty quantum of the observer.  Such a time interval, shorter
than  the  time uncertainty  quantum  of  the observer, can
not be identified. So there must be some uncertainty in
time.  And uncertainty in time will result in uncertainty
in space. Thus EPR paradox turns out to be a good demonstration
of the limitation of the observer. It  is straightforward to see
from (1) that the observer may  think  the  system  has either
infinite light speed or  infinite  space  quantum,   therefore
is  of distance. Actually these two cases are the same in this theory.
When  the  observer measures the spin of one particle, the signal
will  travel  to  the  other particle at a speed larger than light
speed of the observer. You may also say, the distance relative to
the observer between the particles is smaller than the space quantum
of the EPR system. Therefore the two locales of  particles are equal
and indistinguishable in EPR system. This is why the EPR system  is
a whole to the observer.

EPR paradox is a very good example for superlight speed. But in it
lie things which are  more  significant.   Here  we  get  a  deeper
 understanding  for correlation. When we say that a group of particles
 are correlated,  it  means that they are created within one time
 uncertainty  quantum  of  the  observer. Thus we can not decide
 their relative positions in space, just like the  case in EPR
 system. According to modern physics, all matters  in  the
 universe,  especially our physical body,
have a common origin, the originating irregular point of the
Big Bang. This means all the  elementary particles were created
within a time interval so short that we can not identify now.
Thus according to my theory they  were  created  in  one  time
uncertainty  quantum,   and therefore correlated. So on one hand,
we can not decide the distances between these particles in space
in principle.  On  the  other  hand,  everybody correlates with
the world,  and together they form an inseparable whole.
In this way, everybody, in fact every system, establishes
an  unusual corresponding relation between the states
of his body and the universe.  This is something very important
but often ignored in the discussion of EPR paradox. We shall
discuss its profound meaning from a brand-new angle  in
``Quantum Cosmology'',
my third paper of this work.

\noindent {\large\bf IV.     Discussion on the Basis of Statistical Physics}\\

In recent decades, people began to  show  concern  again
for  the  basis  of statistical  physics  because  of
the  achievements  in  nonlinear  sciences. Statistical
physics applies statistics to physics problems, studies
assembly of particles in  a  large  number.   Combining
dynamical  descriptions  with statistics, it tries to
explain the macroscopic properties  of  the  system.
Statistical physics assumes that the macroscopic
properties of the system  is the average of microscopic
counterpart on  
all  possible  microscopic  states under specific macroscopic
conditions. The average is derived with  the  help of the concept
 of statistical assembly,  which is a  collection  of  a  great
 number of imaginary assemblies, independent, with same properties
   and  under same macroscopic conditions. When an isolated system
   reaches equilibrium,  it can be on any  of  the  microscopic
   state  of  the  same  energy  with  same probability. This is
   the so-called ``postulate of equal probability''.

The postulate of  equal  probability,   which  actually  has
other  profound implication  (See the third paper "Quantum
Cosmology" ), got into  continuous controversy as soon as
it was born. Therefore,   Boltzmann  put  forward  the ergodicity
assumption to take place of it. The ergodicity assumption  assumes
that the measure of non-ergodic orbitals in phase  space  of  a
conservative system with limited freedoms are approximately zero.
 Here  ergodic  orbitals are referred to those which linger
in the same area of the energy surface  for the same length of
time. Many people hoped this assumption could  work  as  a good
basis for statistical physics. But it was not as good as expected.
Since the 50s, with the progress in the  research  of  KAM  theorem
 and  related problems, people have gradually come to know that  the
  measure  of  the  KAM constant ring is not zero in many cases, thus
  the  ergodicity  could  not  be realized. But statistical physics,
  which has already made great
achievements, requires some kind of "state-mixing",  which is a stronger
 prerequisite than ergodicity. How to understand the conflict ?   From
 the  reversibility  and certainty of microscopic dynamics to the
 irreversibility and  uncertainty  of statistical physics, what
 causes
the mysterious change ?

Another chronic problem perplexing  statistical  physics  is
the  source  of randomness. During the late scores of years,
advances in chaos  have  changed the outlook of this problem
to a certain extent. Now most people believe that both the
large-number effect  and  the  inner  randomness  in
nonlinear system are all responsible for the randomness
in statistical physics. But what are the natures of  and
relationship  between  the  two  kinds  of randomness?
What is their  relationship  with  the  uncertainty  in
quantum mechanics ? Here, I shall put forward a  picture
based  on  this  theory  of general system to understand
 the problems above.

According to my theory, the uncertainty for a system can
 only  come  from  the outer environment of the system,
 i.e., the finite uncertainty quanta.  This is the intrinsic
  nature of the system. And it  is  this  intrinsic  nature
  that gives rise to all kinds of uncertainty existing in
  forms appropriate  to  the situations. Apparently, the
  higher the synergistic level of the system, the smaller
  the uncertainty in its environment, and thus the less
  likely to be affected by the environment
the system will be.
 You may also  say, its degeneracy is higher in this
 case (degeneracy is the ability of a system to
 correspond to the abundance of its environment,
 see the third paper ``Quantum Cosmology''). Reversely,
 systems of lower synergistic levels, e.g. systems  in
 quantum mechanics, have lower degeneracy and therefore
 larger responding capacity,  thus are more likely to be
 influenced by the environment.   In  such  cases,  because
  of the limitation in our observer systems, we can not get
  to know
 all the influence on
 the system exerted by its environment (e.g. our measurement).
 Therefore it is destined we'll have to  face  uncertainty  and
 randomness.  Generally, observer system has its own inner
 environment and outer environment. Changes invisible to us
 observers might be strong enough  to  cause  notable effect
 in simple systems. Thus comes the uncertainty and randomness.
 So  we can conclude that limitation of the observer is  the
 direct  cause  for  the randomness in quantum mechanics.

As we said above, there is an one-to-one  correspondence
relation  between  the system and its environment. With
this viewpoint we  can  have  an  integrated understanding
 of the randomness in quantum mechanics, statistical physics
 and classical physics. Obviously, for a macroscopic  statistical
  system we'll still have to face randomness as in  quantum  system
  because of our own limitation. But here  with  the  higher
  degeneracy of the system,  it has got some independence and
  we have lost some
information. Thus the increased degeneracy in the system corresponds to the
decreased degeneracy in the observer system, and thus will
add to the  uncertainty  of  the  latter  ( See  the  third  paper  "Quantum
Cosmology" ). When  an  isolated  system  evolves  from  non-equilibrium
to equilibrium, it is divested of its order and so assimilated  by  the
observer system (See the second paper "Where Has Entropy Gone" ). In this
process  of assimilation, the system loses its order or negative entropy, and
at equilibrium  it has
lost its independence and become part of the  observer.   Its  degree  of
 order is lowed down below the limit of the observer, which is revealed by
 the fluctuation in the system. Except for the special case of the perfect
 system, fluctuation is inexorable.

This results in a special state-mixing. In the next paper ``Where Has Entropy
 Gone'',   I'll  show  that  the order of the system and its environment can
 be transformed into each  other,  while their total entropy remains
 constant. In the evolution of  an  isolated system, the observer deprives
 the  system  of  its  order.   This  makes  the observer system more
 ordered or more degenerated, and thus the clock of the observer system
 slower. The  state  of  the  system  that  is  farther  away  from  the
equilibrium has smaller weight thus corresponds to less ordered state of
 the  observer  system,   or  state  with faster clock. Actually, it is
 not very difficult to understand this,   if  we can accept the idea that
  different systems may have different  time  quanta,  thus different
  clocks. Though we haven't got  a  definite  formula  for  the relation
  between the entropy of the system and clock of the observer system,
  it seems that within the above frame  we  can  visualize  that  the
time  of
relaxation is
finite, assembly averages can give  most of the experimental  results
and fluctuations from the average are incidents of small probability.

It is lopsided to consider the problem of   re-establishing  foundation
for statistical  physics  just  as  to  explain  its  probability
feature  with deterministic feature of microscopic dynamics. The
reason is that uncertainty in our environment is inexorable if
our own system has limitation.  Even  if we finally found such
deterministic equations, we  would still be  unable  to grasp
the complete meaning of them because of our own limitation.
Researches in chaos have given good examples in this
aspect. While most of  the  systems in quantum mechanics and
statistical physics are  simple  systems  that  have only
simple synergistic functions,  sophisticated  systems  usually
have  some kinds of nonlinearity for their relatively stronger
synergistic  functions.  The innate randomness of system described
by deterministic equations, or chaos, is an effect in which
nonlinearity  functions.   Commonly,   there  are  two possible
explanations for the origin of probability in a  dynamical  system,
either because we are somewhat ignorant of the initial conditions,
or because of a non-local description that might exist and can take
place  of  orbitals in dynamical system. It is a mistake to think
that  we  will only agree  on  the former because of the feature
of uncertainty quanta. In my opinion,  the  two possibilities are
the same, because  both  of  them  are  resulted  from  the
uncertainty quanta in  the  researcher  ( observer)   system.
It  is  these uncertainty quanta  that  make  us
unable  to  see the  transformation  and unification of space and time.

It should be noted here that although chaos is so popular in
classical  areas that it almost turns to be a fashion,   it
has  not  been  found  in  quantum mechanics yet. This striking
contrast gives a  lot  for  thought.   Obviously, since we have
limitations ourselves, we should find uncertainty no matter  in
classical or quantum systems. Yes, we have. But  uncertainty
gives  different manifestations in the two  kinds  of  systems.
Quantum  mechanics  admitted uncertainty from the very beginning
and takes it as a basic principle,   from which it has developed
its special arithmetical rules. Though such  frankness of
confessing its own limitation is not endearing, it has not
only  put  our understanding for the world a big step forward,
but  also,   to  be  greatly arousing, wiped out the uncertainty
in the form of chaos because of its  open and  well-defined
uncertainty feature.  In  classical  systems, reversely, we
suppose to be omnipotent observer and adopt  mathematics  with
transfinitely
divisible feature. But still we can not  get  rid of uncertainty.
It may appear as the difficulty to grasp a great number,   or as
being incapable to get all initial conditions. Whether we like or
not, the subjective limitation of the observer is destined to be
involved.   Then  why don't we include it in a more advantageous
way  ?   When  you  finally  read through my work, especially
when  you  understand  my  third  paper  "Quantum Cosmology",
you will see that this theory is just  the  right  theory
for the purpose.

Three systems are involved in the studies of chaos,  i.e.
phenomenon  system, theory system and observer system. Each
system has  its  own characteristic synergistic function,
thus there might be some mismatch  among the corresponding
uncertainty quanta. The mismatch in some cases can show the
observer its limitation in its synergistic function.   Take
weather  forecast  as  an example. Here the phenomenon system
is a very complicated one involving  lots of factors and mechanism,
so that  the
contributing  factors  for  different weather may be well within
the outer environment of ordinary people. This  is to say that
there must  be  some  uncertainty  in  the  weather  changes
for ordinary people. Strictly speaking, there  might  be
significant  difference even among mankind (See the second
paper "Where  Has  Entropy  Gone"  ) . The ordinary people
here are referred to those human  observer  system  with
most common cultural and physiological feature. Meteoric
research  apparates  may
change the
synergistic level of the observer system but not our
conclusion.  The theory system in the  research  of
weather  forecast  is  the  group  of integrated differential
equations,  which  are  based  on  the  real  set  in mathematics.
Researches in chaos tell us that long-term  weather  forecasting
is impossible. This is because of the "Butterfly Effect" in
phenomenon system, and because of extreme sensitivity of the
dependence  of  the  mathematical model on initial conditions
and the
transfinite divisibility
in theory system. Suppose our theory system can describe the
phenomenon system well  enough,  then the "Butterfly Effect"
and the sensitivity on initial conditions are  of the same
nature. Both of the two reveal the limitation of the observer
system whose uncertainty feature can not match with the
transfinite divisibility of the theory system. We can not
grasp the delicacy of mathematical model,  just like we can
not grasp the complexity of the environment. In such examples
we see the uncertainty
in our
world through mathematics, the common  intelligence of mankind.
Though the mathematical system has only one uncertainty
quantum(see the following chapter), it is a more sophisticated
system than ordinary people. To get further understanding on
this point, we have to make  some  consideration on mathematics.

\noindent {\large\bf V.      Consideration on Mathematics}\\

One reason that science has deeply rooted in the hearts of
human being is its speculative philosophy and distinctive
quantitative  feature  best  embodied with mathematics. If
the world could really be determined quantitatively from a
few simple rules, we would be really the  master  in  the
world.   In  the nineteenth century and even at the beginning
of  the  twentieth  century  many scientists had such beautiful
dreams. A century has pasted since then.   When we turn round to
review today's science, we 
find regretfully that  there  are very few areas in the varieties
of natural sciences in which such dreams have come true. Even in
physics, which is the underlying subject for other sciences, such
dreams have turned out to be soap bubbles. Quantum  mechanics  was
the greatest achievement that the science of the twentieth century
could be  proud of. With  its  brand-new  characters  like  the
uncertainty  principle  and probability explanation of  wave
function,   it  perplexed  many
idealistic physicists. It had been not long  since  people
generally  accepted  quantum mechanics, when another magician came on
the  stage  quietly.   That  is  the nonlinear science. It
has brought
about a lot of new achievements and given a big shock to old ideas.
Although  both  quantum  mechanics  and  nonlinear science have given
deadly blows to the dream to determine the world from only a few simple
rules, the latter seems more complicated so  that  some  people think
the chaotic delicacy is
beyond human
comprehension. In my opinion, this is because that the mathematics
plays a special role in it.

The  original and direct source of mathematics was human experience.
Man is the most popular and important system. Every man  has  his  own
Inner  and  outer  environment,   his  own uncertainty quanta. This is
why  the  natural  set,   with  apparent  quantum character, was the
simplest and therefore the first mathematics to come  into being. The
natural set is in accordance with individual experience and thus a
system simpler than individual human system. But with the  efforts
of  many
mathematicians, mathematics turns to be, it seems, an independent
system which goes further and further beyond individual experience.
In this system, there are the infinite and  the  infinitesimal,
infinitely  divisible  real  set,  transfinite ordinals and cardinals
and so on. Even  if  the  significance  of these work is not completely
clear, it should not be refuted because  it  has brought about abundance
of ideological fruits for human being. Because of the development of the
linkage between human
individuals, mathematics today  has long become a large system parallel
to the whole human being. It has gained a certain independence, occupied
more and more domains,   and  tried  to  find advancing motivity from
within its own consistency. Such trend culminated  in the wave of
axiomization  of  the  twentieth  century,   during  which  formal
descriptions were too highly  praised,   the  ties  between  mathematics
and physics was weakened, and practical source of mathematics was
concealed.
It seems that mathematics
has gained some vitality from  human  being  and  also motivity
for its
own development. Unfortunately the mansion of formal
mathematics is not  perfect  and still haunted
by randomness. This reminds us that there is  still
limitation in today's mathematics or in the way we
understand  it.   Today  mankind  has become a closely
related whole and almost every individual has some character
of the whole human culture to some extent. There may be no
infinite  in  the experience
of an individual, but
human  being can  accept  it  because  the experience and therefore
the  intelligence  of  the  whole  human  being
are infinite by themselves. Though science and culture
today are truly to be proud of, they are still
systems with limitations. So it is with mathematics.
It needs to get new vitality from the innovation of physics ideas.

Most of the mathematics we use today are based on the
real set, which has the property of transfinite divisibility
that is contradictory  to  the  quantum feature of human individual
system, so there must be some results beyond
our comprehension. In linear mathematics
people  can  always  presume  that  the change of a variable is in
quantum (no matter how small the quantum is), which is in  accordance
with  our  experience.   But  nonlinear  calculation  has completely
lost
the familiar feature
of quantum. It is iterative, accumulative and divisible. The transfinite
divisibility of the real set is a mathematical idealization.  Not only
is it different from experience of human individual,  but also it is
related in a  profound  way  to  many  famous  problems  which
symbolize the mystique of the nature of mathematics. In fact,
most  systems have uncertainty in their synergistic functions
so that they can not fix  its environment  with  infinite  accuracy.
Thus  very  perhaps
the
infinitely divisible delicacy of mathematics has complicated many
problems, thus made it incapable of providing clear insights. This
is why I think it necessary  to promote the research  for  a  new
kind  of  mathematics  characterized  with uncertainty quantum. In
this new mathematics there will be new definition for limit and new
algorithm. The infinite and infinitesimal,   which  have  been strange
to each other in present mathematics, will be linked  together  in  a
wonderful and profound relation.
I believe such mathematics will reveal a new
scene of the world for human being.

Apparently, just as quantum
mechanics admitted  its  uncertainty  openly  and frankly, the
new mathematics should have certain fuzzy feature. We know  that
the classical mathematics, which is based on ZFC axiom set theory,
does  not cope with fuzzy objects because of its clarity in the way
it  constructs  its sets. So the set theory in common sense can not
form the basis  for  the  new mathematics. Though the present fuzzy
mathematics  has  obtained  a  certain fuzzy
feature with the introduction of the concept of membership,
there  are still two points in its nature which shackle
it well  within  the  domain  of classical
mathematics. That is the transfinite divisibility and  the  law  of
excluded middle. The present fuzzy mathematics gives no criticism to
the  law of excluded middle. And its logic, multi-value logic, is
still in  the  sense of classical mathematics. As we pointed above,
the  transfinite  divisibility is not in accordance with the reality.
It is
not difficult to visualize  that the law of  excluded middle is incorrect
whenever and wherever the transfinite divisibility fails. The reason is
that the empty set in  such  case  is  not really empty. There  is  an
axiom for empty set in the ZFC system which says that there exists  an
empty set that contains no element. In fact, such empty set is available
only in the real set which has the character of  transfinite  divisibility
suitable  for describing ideal case rather than reality. When the
transfinite  divisibility is futile, there must be a smallest unit
( uncertainty  quantum) .   If  such smallest unit is taken to be
the empty set,  then  there  must  be  structure within the empty
set which we can not know by nature. And because we can  not know,
we can not exclude the possibility that there  might  be  infinite
and transfinite structure in it. That is, we can not exclude the
possibility  that  there  are  divergent  structures inside the
empty set. As a matter of fact such cases
have already been seen in fractal geometry. On the other hand,
according  to  the  definition  of  the empty set, because we
can know nothing about the structure beyond the infinite,  all
structure beyond the infinite is also related with the empty
set.  Thus the infinite and the infinitesimal will integrate
harmonically in the new mathematics. The classical mathematics,
with its  transfinite  divisibility feature, will turn to be a
special case in the new mathematics
when the empty set is really empty. Thus I
believe, the first step to construct the new set theory, which
will serve as the basis for new mathematics,   is  to
find  a reasonable definition for empty set.

Of course, nobody can deny the great achievements  of  classical
mathematics. What I give here is only the most tentative idea
about the new  mathematics.  It  will  have  non-local,
integrated  and  systematic  character.    Our understanding
for the world will be greatly deepened with the development
of the new mathematics. Like in quantum mechanics, we will
not lose quantitative advantages because of the fuzzy feature
of the
new mathematics,
but will we can have a panorama of the uncertainty that we can
not see with classical mathematics. Thus as a matter of fact we
will have more certainty than before in a broader sense.  On
the other hand, once we get all the qualitative conclusions we
need, why  do  we still need quantitative description ?
Quantitative or qualitative, this  may be the ideological
difference between local and integrated mathematics.

Here we shall  also  put  forward  a  picture  to  understand
the  continuum hypothesis (referred to as CH in the following).
In  1878, Cator  conjectured that the weight of the continuum was
the second  transfinite  cardinal.   But until his death forty
years later, the great  mathematician  did  not  give  a proof
for the conjecture. That is the well-known continuum hypothesis.
In 1900, the great mathematician Hilbert listed in a report
twenty-three  outstanding difficult mathematical problems, the
first of which was to prove the CH.  But the proof here
degenerated into such requirement as to show  the  problem
is unsolvable in the given sense under consideration. After
the  work  of  Gödel and Kohn, many mathematicians realized
that the CH  was  undecidable  in  the common axioms of set
theory, the difficulty might be not purely  mathematical and
the solution would depend on innovations in the basis of
mathematics.  In our theory, the weight of a set is comparable
to the abundance
of environment for
a general system (See the second paper "Where Has Entropy Gone" ).
As I showed above, the real set is transfinitely divisible. Because
of  this property, the null is a real empty set without any
structure.   But  in  ZFC system and also other formal axiom
systems,   there  are  surely  independent proposition. So
they are systems with outer environment, their empty sets
are not really empty. Therefore compared with ZFC system,
the  continuum  is  a more sophisticated  system  with  higher
synergistic  level.   Obviously,   in principle
a  low-levelled  system  can  not  understand,   let
alone  prove,  propositions in a  higher-levelled  system,
because  the  former  has  more uncertainty. So we see that
the reason that no solution for the CH  has  been found
yet is indeed having been under  insufficient  prerequisites
or in an incorrect sense. But the effort  in  solving  the
problem  as  well  as  the thinking aroused in the process
is a very splendid
page  in  the  history  of
human ideology. It still gives us inspiration today.
\end{document}